\documentstyle[epsfig]{mn}

\newif\ifAMStwofonts

\def\gtsima{$\; \buildrel > \over \sim \;$}
\def\ltsima{$\; \buildrel < \over \sim \;$}
\def\gsim{\lower.5ex\hbox{\gtsima}}
\def\lsim{\lower.5ex\hbox{\ltsima}}

\def\Msun{{M_\odot}}
\def\Zsun{{Z_\odot}}
\def\be{\begin{equation}}
\def\ee{\end{equation}}

\ifoldfss
  \ifCUPmtlplainloaded \else
    \NewTextAlphabet{textbfit} {cmbxti10} {}
    \NewTextAlphabet{textbfss} {cmssbx10} {}
    \NewMathAlphabet{mathbfit} {cmbxti10} {} 
    \NewMathAlphabet{mathbfss} {cmssbx10} {} 
  \fi
  \ifAMStwofonts
    \ifCUPmtlplainloaded \else
      \NewSymbolFont{upmath} {eurm10}
      \NewSymbolFont{AMSa} {msam10}
      \NewMathSymbol{\upi}     {0}{upmath}{19}
      \NewMathSymbol{\umu}     {0}{upmath}{16}
      \NewMathSymbol{\upartial}{0}{upmath}{40}
      \NewMathSymbol{\leqslant}{3}{AMSa}{36}
      \NewMathSymbol{\geqslant}{3}{AMSa}{3E}

      \let\leq=\leqslant 
      \let\geq=\geqslant \let\ge=\geqslant
    \fi
  \fi
\fi 

\ifnfssone
  \newmathalphabet{\mathit}
  \addtoversion{normal}{\mathit}{cmr}{m}{it}
  \addtoversion{bold}{\mathit}{cmr}{bx}{it}
  \newmathalphabet{\mathbfit} 
  \addtoversion{normal}{\mathbfit}{cmr}{bx}{it}
  \addtoversion{bold}{\mathbfit}{cmr}{bx}{it}
  \newmathalphabet{\mathbfss} 
  \addtoversion{normal}{\mathbfss}{cmss}{bx}{n}
  \addtoversion{bold}{\mathbfss}{cmss}{bx}{n}
  \ifAMStwofonts
    \ifCUPmtlplainloaded \else
      \UseAMStwoboldmath
      \makeatletter
      \new@mathgroup\upmath@group
      \define@mathgroup\mv@normal\upmath@group{eur}{m}{n}
      \define@mathgroup\mv@bold\upmath@group{eur}{b}{n}
      \edef\UPM{\hexnumber\upmath@group}
      \new@mathgroup\amsa@group
      \define@mathgroup\mv@normal\amsa@group{msa}{m}{n}
      \define@mathgroup\mv@bold\amsa@group{msa}{m}{n}
      \edef\AMSa{\hexnumber\amsa@group}
      \makeatother
      \mathchardef\upi="0\UPM19
      \mathchardef\umu="0\UPM16
      \mathchardef\upartial="0\UPM40
      \mathchardef\leqslant="3\AMSa36
      \mathchardef\geqslant="3\AMSa3E

      \let\leq=\leqslant 
      \let\geq=\geqslant \let\ge=\geqslant
    \fi
  \fi
\fi 

\ifnfsstwo
  \DeclareMathAlphabet{\mathbfit}{OT1}{cmr}{bx}{it}
  \SetMathAlphabet\mathbfit{bold}{OT1}{cmr}{bx}{it}
  \DeclareMathAlphabet{\mathbfss}{OT1}{cmss}{bx}{n}
  \SetMathAlphabet\mathbfss{bold}{OT1}{cmss}{bx}{n}
  \ifAMStwofonts
    \ifCUPmtlplainloaded \else
      \DeclareSymbolFont{UPM}{U}{eur}{m}{n}
      \SetSymbolFont{UPM}{bold}{U}{eur}{b}{n}
      \DeclareSymbolFont{AMSa}{U}{msa}{m}{n}
      \DeclareMathSymbol{\upi}{0}{UPM}{"19}
      \DeclareMathSymbol{\umu}{0}{UPM}{"16}
      \DeclareMathSymbol{\upartial}{0}{UPM}{"40}
      \DeclareMathSymbol{\leqslant}{3}{AMSa}{"36}
      \DeclareMathSymbol{\geqslant}{3}{AMSa}{"3E}

      \let\leq=\leqslant 
      \let\geq=\geqslant \let\ge=\geqslant
    \fi
  \fi
\fi 

\ifCUPmtlplainloaded \else
  \ifAMStwofonts \else 
    \def\upi{\pi}
    \def\umu{\mu}
    \def\upartial{\partial}
  \fi
\fi

\title{Mining the Galactic Halo for Very Metal-Poor Stars}
\author[Salvadori, Ferrara, Schneider, Scannapieco \& Kawata]
{S. Salvadori$^{1}$, A. Ferrara$^{2}$, R. Schneider$^{3}$, E. 
Scannapieco$^{4}$ \& D. Kawata$^{5}$\\
$^1$SISSA/International School for Advanced Studies, Via Beirut 4, 
34100 Trieste, Italy\\ 
$^2$Scuola Normale Superiore, Piazza dei Cavalieri 7, 56126 Pisa, Italy\\
$^3$INAF/Osservatorio Astrofisico di Arcetri, Largo Enrico Fermi 5, 
50125 Firenze, Italy\\
$^4$School of Earth and Space Exploration, Arizona State University, P.O. 
Box 8714, Tempe, AZ, 85287-1404\\
$^5$Mullard Space Science Laboratory, University College London, 
Holmbury St. Mary, Dorking, Surrey, RH5 6NT\\}
\date{}

\pagerange{\pageref{firstpage}--\pageref{lastpage}}
\pubyear{}

\begin{document}

\maketitle 
\label{firstpage}

\begin{abstract}
We study the age and metallicity distribution function (MDF) of metal-poor 
stars in the Milky Way halo as a function of galactocentric radius by 
combining N-body simulations and semi-analytical methods. We find that 
the oldest stars populate the innermost region, while extremely metal-poor 
stars are more concentrated within $r<60$~kpc.
The MDF of [Fe/H]$\leq -2$ stars varies only very weakly 
within the central $50$~kpc, while the relative contribution of 
[Fe/H]$\leq -2$ stars strongly increases with $r$, varying from $16\%$ 
within $7$~kpc~$<r<20$~kpc up to $\geq 40\%$ for $r>20$~kpc. This is due 
to the faster descent of the spatial distribution (as seen from Earth) 
of the more enriched population. This implies that the outer halo 
$<40$~kpc is the best region to search for very metal-poor stars.
Beyond $\sim 60$~kpc the density of [Fe/H]$\leq -2$ stars is maximum 
within dwarf galaxies. All these features are imprinted by a combination of 
(i) the virialization epoch of the star-forming haloes, and 
(ii) the metal enrichment history of the Milky Way environment.  
\end{abstract}


\section{Introduction}
Very metal-poor stars ([Fe/H]$\leq -2$) represent the living fossils of the 
first stellar generations. Their observation is crucial, as they may provide 
fundamental insights on both the properties of the first stars and on the 
physical mechanisms governing the early stages of galaxy formation, such as 
feedback processes. An intrinsic problem that observers have to face is that 
old, very metal-poor stars, are extremely rare in the solar neighborhood, 
comprising no more that the $\sim 0.1\%$ of the stars within a few kpc of the 
Sun (Beers et~al. 2005). 

During the past years several surveys focused on such elusive stellar
populations, both in the Milky Way (MW) halo and in nearby dwarf satellites, 
providing an increasing amount of data. At the moment, the metallicity 
distribution function (MDF) of Galactic halo stars (Beers et~al. 2005)
represents one of the most important observational constraints. Indeed, 
it consists of $2756$ halo field stars observed within $\lsim 20$~kpc 
of the Sun (Beers \& Christlieb 2005), covering a huge metallicity range 
which spans from [Fe/H]$=-2$ down to [Fe/H]$=-4$. Despite such a large 
sample, the number ($\approx 300$) of extremely metal-poor stars ([Fe/H]$<-3$) 
turns out to be still insufficient to put solid constraints on the properties 
of the first stars (Tumlinson 2006, Salvadori, Schneider \& Ferrara 2007), 
as the MDF is a rapidly increasing function of [Fe/H]. Hence, understanding 
where these stars are preferentially located is an urgent theoretical 
question.  

Carollo et~al. (2007) have recently done an accurate kinematic study of  
$\sim 10.000$ calibration stars of the SDSS Data Release 5, finding that 
the ``outer'' halo, $r \ge 15$~kpc, includes a larger fraction of [Fe/H]$<-2$ 
stars and peaks at lower metallicity than the ``inner'' halo ($r<15$~kpc). 
This evidence poses challenging questions about the physical origin of such 
segregation and the variation of the MDF with galactocentric radius. 

In this study we investigate the spatial distribution of metal-poor halo stars 
by combining an high-resolution N-body simulation for the formation of the MW 
(Scannapieco et~al. 2007), with a semi-analytical model (Salvadori, Schneider 
\& Ferrara 2007, hereafter SFS07) that follows the stellar population history 
and the chemical enrichment of the Galaxy along its hierarchical tree, 
successfully reproducing several properties of the MW and its dwarf satellites 
(Salvadori, Ferrara \& Schneider 2008, hereafter SFS08; Salvadori \& Ferrara 
2009).
\section{Summary of the model}
\subsection{The N-body simulation}
We briefly summarize the main features of the N-body simulation referring to 
Scannapieco et~al. (2007), and references therein, for a detailed description. 
We simulate\footnote{We adopt a $\Lambda$CDM cosmological model with $h=0.71$, 
$\Omega_{0}h^2=0.135$, $\Omega_{\Lambda} = 1 - \Omega_0$, 
$\Omega_{b}h^{2}=0.0224$, $n=1$ and $\sigma_8 = 0.9$.} a MW-analog galaxy with 
the GCD+ code (Kawata \& Gibson 2003a) using a multi-resolution technique 
(Kawata \& Gibson 2003b) to achieve high resolution in the regions of interest. The initial conditions at $z=56$ are constructed using the public software 
GRAFIC2 (Bertschinger 2001). The highest resolution region is a sphere with 
a radius 4 times the virial radius of the system (i.e. the MW) at $z=0$, the 
dark matter (DM) particles mass and softening length are respectively 
$7.8\times 10^5\Msun$ and $540$~pc. The system consists on about $10^6$ 
particles within $r_{vir}$; its virial mass and radius are respectively 
$M_{vir}=7.7\times 10^{11}\Msun$ and $r_{vir}=239$~kpc, roughly consistent with 
the observational estimates ($M_{vir}=10^{12} \Msun$, $r_{vir}=258$~kpc) of 
the MW (Battaglia et~al. 2005). The simulation data is output every $22$~Myr 
between $z=8-17$ and every $110$~Myr for $z<8$. At each output a 
friend-of-friend group finder is used to identify the virialized DM haloes by 
assuming a linking parameter $b=0.15$ and a threshold number of particles of 
$50$. A low-resolution simulation including gas physics and star formation (SF)
has been used in order to confirm that the initial conditions will lead to a 
disk formation. While gas physics is essential to reproduce the spatial 
distribution of disk stars, an N-body approach is suitable to investigate the 
halo (and bulge) population we are mostly interested in. 
\subsection{The semi-analytical model}
We briefly describe the basic features of the model implemented in the code 
GAMETE (GAlaxy MErger Tree \& Evolution) referring the reader to SSF07 and 
SFS08 for more details. We trace the evolution of gas and stars inside each 
halo of the hierarchy (group of DM particles) by assuming the following 
hypotheses: (a) at the highest redshift of the merger tree, $z=17$, the gas 
has a  primordial composition; (b) stars can only form in haloes of mass 
$M_h > M_4(z)=3\times 10^8\Msun (1+z)^{-3/2}$ 
($M_h > M_{30}(z)=2.89\times M_4(z)$) prior to (after) reionization, here 
assumed to be complete at $z=6$; (c) in each halo the SF rate is proportional 
to the mass of cold gas; (d) according to the critical metallicity scenario 
(Schneider et~al. 2002, 2006) low-mass stars with a Larson Initial Mass 
Function form when the gas metallicity $Z \ge Z_{\rm cr}=10^{-5\pm 1}\Zsun$; 
for $Z < Z_{\rm cr}$ massive Pop~III stars form with a reference mass 
$m_{\rm PopIII}=200\Msun$.

\begin{figure}
  \centerline{\psfig{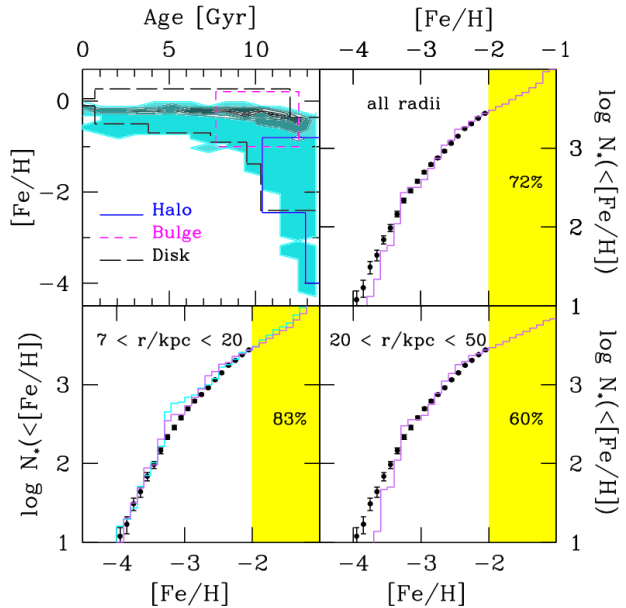}}
  \caption{{\it Left upper panel}: comparison between the observed 
    (rectangles) and simulated (shaded regions) age-metallicity distribution 
    of MW stars. Rectangles show the observed relation for different Galactic 
    components (Freeman \& Bland-Hawthorn 2002): the halo (blue solid 
    rectangles); the thin and thick disk (black long-dashed rectangles); the 
    bulge (violet short-dashed rectangles). The colored shaded areas correspond 
    to regions that include, from the darkest to the lightest, the 
    $(30,62,90,100)\%$ of the total number of relic stars produced in the 
    simulation, $M^*_{tot}\approx 4\times 10^{10}\Msun$. 
    {\it Top upper and lower panels}: 
    comparison between the cumulative MDF observed in the Galactic halo within 
    $\lsim 20$~kpc of the Sun (points with Poissonian error bars, Beers et~al. 
    2005) and those produced in the simulation at different radii (violet 
    histograms) normalized to the number of observed stars. The cyan histogram 
    shows the $7<(r/{\rm kpc}) <20$ MDF for the inhomogeneous mixing case. 
    For each range of radii the numbers show the percentage of 
    $-2<$[Fe/H]$\leq -1$ stars (shaded area) with respect to the total number 
    of [Fe/H]$\leq -1$ stars.}
  \label{fig:1}
\end{figure}
We describe the enrichment of gas within proto-Galactic haloes and diffused in 
the MW environment, or Galactic Medium (GM), by including a simple description 
of supernova (SN) feedback. Metals and gas are assumed to be instantaneously 
and homogeneously mixed with the gas (implications discussed in SSF07); we 
assume the Instantaneous Recycling Approximation (IRA, Tinsley 1980). At each 
time-step the mass of gas, metals and stars within each halo is equally 
distributed among all its DM particles, and such information propagated to the  
next integration step. The same procedure is applied to metals ejected into the 
GM; as a result, the chemical composition of newly virializing haloes depends 
on the enrichment level of the GM out of which they form. Finally, the 
properties of long-living metal-poor stars hosted by each DM particle are 
stored to recover their spatial distribution at $z=0$.

As a rough estimate of the impact of the perfect mixing assumption above we 
have also explored a simple case of inhomogeneous metal mixing. The latter 
is modeled by computing the instantaneous filling factor 
$Q=(\sum_i 4\pi R^3_b(i)/3V_{MW}(z))$ of the metal bubbles inside the critical 
MW volume $V_{MW}(z)=30 (1+z)^{-3}$~Mpc$^3$ and by randomly enriching a 
fraction $F=1-exp(-Q)$ of GM particles. Note that this only provides an upper 
limit on F, as a clustered  system, such as the stars we are considering, will 
have a smaller filling factor. A simple Sedov-Taylor blastwave solution 
(Sec.~3.6 of SFS07) is used to estimate the bubble radii $R_b(i)$. This case 
is compared with the perfectly mixed one in Fig.~1 (lower panel on the left).
\section{Results}
The model is calibrated, i.e. the SF and SN wind efficiencies are fixed, by 
simultaneously reproducing the global properties of the MW (stellar/gas mass 
and metallicity) and the Galactic halo MDF as in SSF07.
\subsection{The age-metallicity relation}
A first test of our model results is a comparison with the observed stellar 
age-metallicity; the results are shown in the upper panel of Fig.~1. 
The simulated distribution can be virtually divided into three main regions 
defined by the prevailing formation mode of the stars contained within them. 

\begin{figure}
  \centerline{\psfig{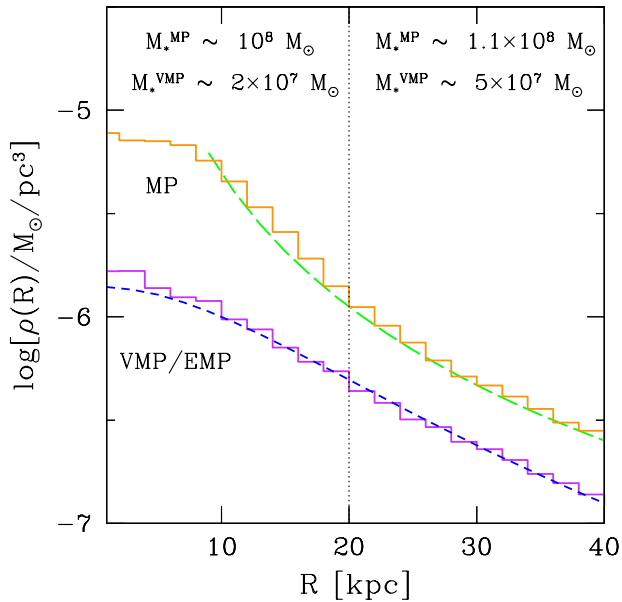}}
  \caption{Average density profile of $-2<$[Fe/H]$<-1$ (top orange histogram) 
    and [Fe/H]$\leq -2$ stars (bottom violet histogram) as a function of the 
    distance from the Earth, $R$. Short- (long-) dashed curve is the 
    $\beta$-model $(1+(R/R_c)^2)^{-3\beta/2}$ (power-law $R^{-\gamma}$) 
    best-fit to the stellar distribution.}
 \label{fig:2}
\end{figure}
The first region (age $> 13$ Gyr, or $z>7.5$) is populated by old stars 
covering almost the entire metallicity range, from [Fe/H]$\sim -0.3$ to 
[Fe/H]$=-4.2$, and which correspond to the observed {\it bulge} and {\it halo} 
components (see the rectangles in the panel). These stars formed in 
proto-galactic haloes associated to high ($> 2\sigma$) density fluctuations, 
that virialized during the early stages of Galaxy formation at $z>7.5$. These 
first stellar generations enriched the interstellar medium (ISM) of their host 
galaxies up to [Fe/H]$>-2$ (a process we dub as {\it self-enrichment}), 
quenching the 
formation of additional very metal-poor stars. At the same time, metals are 
expelled by SN feedback in the GM, thus increasing its metallicity above 
$Z_{\rm cr}$ by $z=11$ and allowing long-living metal-poor stars to form in 
newly virialized haloes {\it accreting} their gas from the GM. From $z=11$ to 
$z=7.5$ coeval formation of metal-poor and metal-rich stars occurs in different 
objects through accretion and self-enrichment processes. During this epoch 
$-3<$[Fe/H]$<-2$ stars are produced via merging of self-enriched and accreted 
haloes. Unlike in Scannapieco et al. (2007), no [Fe/H] $\leq  -4$ 
stars form below $z=7$ because of the GM mixing approximation assumed here.

The second region (Age $< 13$ Gyr, [Fe/H]$<-0.3$) is filled by stars which 
almost span the entire range of ages and metallicities along a relation on 
which the iron-abundance increases with decreasing age ({\it halo, bulge, thick 
and thin disk}). These stars formed in accreting haloes; the minimum [Fe/H] 
value of the stellar distribution at different epochs reflects the iron 
evolution of the MW environment. Stars located in this area only represent 
$< 10\%$ of the total stellar mass at $z=0$.

The third region ([Fe/H]$>-0.3$) is populated by iron-rich stars formed in 
self-enriched haloes, which therefore span all the possible ages ({\it bulge, 
thin and thick disk}). The bulk of the stellar mass resides in this region 
and corresponds to the broad peak of the SF rate ($2 < z <5$, see for a 
representative SF history Fig. 1 of Evoli, Salvadori \& Ferrara 2008). Note 
that the most iron-rich stars in the simulation with ages $<12$~Gyr, are 
poorer in iron than those observed. This systematic effect may be a consequence 
of neglecting the contribution of SNIa to gas enrichment. However because this 
study is mostly concentrated on the spatial distribution of {\it old} and 
{\it iron-poor} Galactic halo stars this does not affect the main results. 
Even if we consider the possible existence of a ``prompt'' SNIa component, 
with lifetime of SNIa of $0.1$~Gyr (Mannucci, Della Valle \& Panagia, 2006),
this is typically longer than the evolutionary time-scale of SNII ($<0.03$~Gyr).
Therefore, the iron produced by the first SNIa will only marginally contribute 
to pollute an ISM which has been already largely pre-enriched by {\it several} 
generations of SNII and Pop~III stars.

In the following we will focus on the properties of old metal-poor 
[Fe/H]$<-1$ stars, whose features are unaffected by the lack of disk 
formation and SNIa contribution of our study. By excluding from our sample
all the stars residing at distances $<1$~kpc from the Galactic plane, 
we remove possible contamination by thin/thick disk stars.
\subsection{Metallicity distribution}
In the left lower panel of Fig.~1 the Galactic halo MDF observed by Beers 
et~al. (2005) is compared with the simulated one at galactocentric distances 
$7<(r/{\rm kpc}) <20$. The agreement between the model results for a radial 
interval representative of the observed region and the data is very good. 
In the same panel we show for comparison the simulated $7<(r/{\rm kpc}) <20$ 
MDF for the inhomogeneous mixing case (cyan histogram). The two mixing 
prescriptions only yield a marginal difference, mostly concentrated in the 
range $-3.5<$[Fe/H]$<-3$ (see Sec.~4 for the discussion). These results 
encourage us to formulate explicit predictions for the radial dependence 
of the simulated MDF. To explore this point, we compare the {\it same} 
observational data with theoretical MDFs obtained in {\it different} radial 
bins and normalized to the number of observed stars. We note that the all-radii 
MDF (left upper panel) already provides a satisfactory match of the data 
implying that the observational sample obtained in the above $r$-range provides 
a good proxy for the all-radii one. A marginal discrepancy at the low-Fe end of 
the distribution is found when comparing the observed data with theoretical 
MDFs derived for $r>20$~kpc (lower panels), which becomes more sensible as $r$ 
increases. In conclusion, we find that the MDF varies only very weakly with 
radius. On the contrary, and in agreement with recent findings by Carollo 
et~al. (2007)
and De Lucia \& Helmi (2008), the relative contribution of $-2<$[Fe/H]$< -1$ 
stars to the total MDF strongly depends on $r$, varying from $84\%$ within 
$7$~kpc$<r<20$~kpc, down to $\leq 60\%$ for $r>20$~kpc.
\begin{figure*}
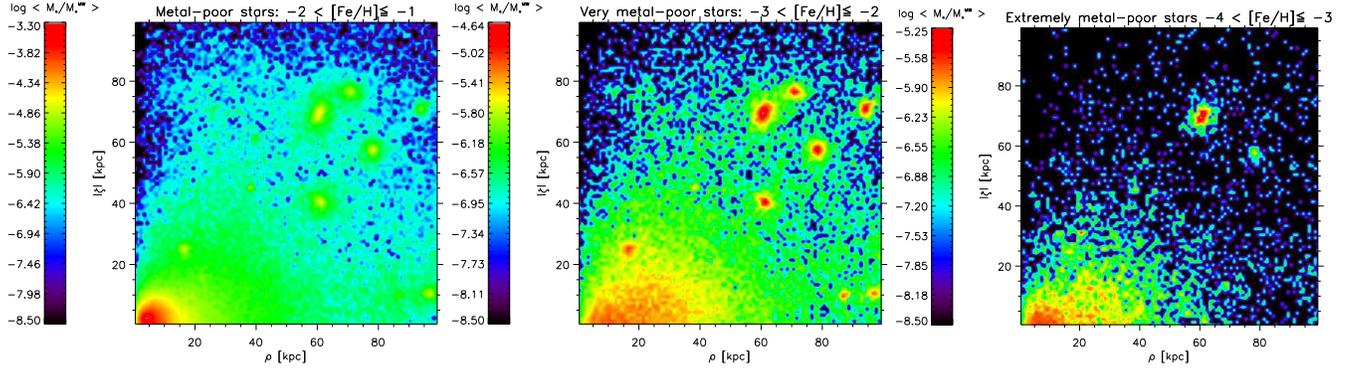

  \centerline{\psfig{figure=FIG3a.epsi,width=5.80cm,angle=0}
    \psfig{figure=FIG3b.epsi,width=5.80cm,angle=0}
    \psfig{figure=FIG3c.epsi,width=5.80cm,angle=0}}
  \caption{Mass distribution of metal-poor (MP) $-2<$[Fe/H]$\leq -1$ (left 
	panel), very metal-poor (VMP) $-3<$[Fe/H]$\leq -2$ (middle panel), 
	and extremely metal-poor (EMP) $-4<$[Fe/H]$\leq -3$ (right panel) 
	stars, in the cylindrical coordinate plane ($r,|\zeta|$), normalized 
	to the total MW stellar mass in the simulation 
	$M^*_{tot}\approx 4\times 10^{10}\Msun$.}
\label{fig:3}
\end{figure*} 

To better understand these features it is helpful to analyze Fig.~3, which 
offers a spatial visualization of the MDFs in the $\rho-\zeta$ cylindrical 
coordinate plane, where $z$ is the rotation axis and $r^2={\rho}^2+{\zeta}^2$. 
We show the results for the central $100\times 100$~kpc$^2$ region. Each panel 
of the figure displays a subset of relic stars in a different metallicity 
range: {\it metal-poor} (MP), $-2<$[Fe/H]$\leq -1$, {\it very metal-poor} 
(VMP), $-3<$[Fe/H]$\leq -2$, {\it extremely metal-poor} (EMP), 
$-4<$[Fe/H]$\leq -3$. The colors show the total mass of stars contained in an 
annulus of radial width within $1$~kpc normalized to the total MW stellar mass. 

The stellar distribution closely follows the dark matter one, i.e. it is denser 
towards the center and in the 10 dwarf galaxies found in $50$~kpc~$<r<100$~kpc. 
The radial dependence of the MPs distribution (left panel) is very steep, 
varying by more than 2 orders of magnitude in the inner $50$~kpc; into the 
same region instead VMP/EMP stars (middle/right panels) are much more uniformly 
distributed and exhibit a central core. 
This is further illustrated in Fig.~2, which shows the average density 
profiles of MP and VMP/EMP stars as a function of the distance from 
the Earth, $R$. While for $R>10$~kpc the density of MP stars closely 
follows a power-law in radius, $R^{-2.2}$, that of VMP/EMP stars is well 
approximated by a $\beta$-function, $[1+(R/R_c)^2]^{-3\beta/2}$, with 
$\beta=1.4$ and $R_c=20$~kpc. It follows that the relative contribution 
to the MDF of more pristine stellar generations becomes gradually more 
important at large distances (Fig.~1). 
Beyond $r\sim 60$~kpc very metal-poor stars are mostly concentrated within 
dwarf satellites, which are clearly identified in Fig.~3. This is in agreement 
with well-known evidence that the MDF in dwarf galaxies is shifted towards 
lower-[Fe/H] with respect to the Galactic one (Helmi et~al. 2006). 
Interestingly extremely metal-poor stars are found only in two dwarfs and, even 
in these objects, they represent a sub-dominant stellar population 
($\leq 13\%$). Beyond 60~kpc the number of EMPs drops implying that this 
population is more condensed within such a region.

What determines the spatial distribution of stars with different [Fe/H]?
In addition to the underlying structure formation governed by DM, there are 
two key ingredients: (i) the virialization epoch of the star-forming haloes, 
which affects the final distribution of DM and hence of stars; (ii) the metal 
enrichment history of the GM, setting the initial Fe-abundance of the ISM in 
newly virializing haloes. In Fig.~4 we show the average formation redshift of 
DM particles hosting [Fe/H]$<-1$ stars, $\langle z\rangle$, in the $\rho-\zeta$ 
plane. The oldest stars populate the innermost region; moreover, 
$\langle z\rangle$ gradually decreases with $r=\sqrt{\rho^2+\zeta^2}$. 
Beyond $\sim 30$~kpc on average $\langle z\rangle < 7$. As for $z < 8$ the GM 
has been already enriched up to [Fe/H]$_{GM} \approx -3$, extremely metal-poor 
stars become more rare in such outer regions; this explain their spatial 
condensation. Very metal-poor stars, instead, extend up to 100~kpc as 
[Fe/H]$_{GM}\approx -2$ when $z=5$. Finally, as metal-poor stars predominantly 
form via self-enrichment their spatial distribution is unaffected by the GM 
enrichment and it is solely determined by hierarchical history of collapsed 
structures.

A final remark concerns dwarf satellite galaxies. Beyond $\sim 30$~kpc the 
dwarf systems can be identified as clumps of high $\langle z\rangle$ against 
the more uniform background. All satellites found in the simulation are 
``classical'' dwarf galaxies, i.e. they have $L>10^5 L_{\odot}$. Only two of 
them, corresponding to rare $>2\sigma$ fluctuations, were hosting Pop~III stars 
as they virialized and began to form stars when $z>11$ and $Z_{GM}<Z_{cr}$.
The powerful explosions following the evolution of Pop~III 
stars\footnote{Massive Pop~III stars evolve as pair-instability SN. For 
$m_{\rm PopIII}=200\Msun$ the explosion energy is $2.7\times 10^{52}$~erg} 
caused the complete blow-away of gas and metals; 
long-living stars only form at later times when more {\it pre-enriched} gas 
is collected by the dwarfs through accretion and merging processes. No clear 
imprint of their pristine formation can be found in these galaxies, which have 
similar stellar populations ($\langle$[Fe/H]$\rangle \sim -2$) of ``normal'' 
dwarf satellites. Note however that these galaxies represent the most massive 
dwarfs we found with $M_h=1-2\times 10^9 \Msun$ and 
$M_*=0.7-1.2\times 10^7 \Msun$. The remaining $80$\% of dwarfs correspond to 
$<2\sigma$ fluctuations which virialize at later epochs $z=(6-8)$ when 
[Fe/H]$_{GM}>-3$. The lack of [Fe/H]$<-3$ stars (Helmi et~al. 2006) is 
hence naturally explained in these objects which have a dark matter 
$M_h = (1-7)\times 10^8\Msun$ and stellar mass content 
$M_* = (0.5-7.5) \times 10^6 \Msun$ consistent with that of the observed 
dwarf spheroidal galaxies.
\section{Discussion}
Old, [Fe/H]$<-2$ stars, are intrinsically rare in the Galaxy, representing 
only $\leq 1\%$ (i.e. $\leq 5\times 10^8\Msun$) of the total stellar mass. 
This makes the selection of VMP stars one of the major challenges of stellar 
surveys dovoted to their investigation. Our study shows that: 
(i) the density of $-2<$[Fe/H]$<-1$ stars as a function of distance from 
Earth is very steep, following a power-law, $R^{-\gamma}$, with $\gamma=2.2$; 
on the contrary (ii) the density distribution of VMP/EMP stars exhibits a 
central core, closely following a $\beta$-function, 
$[(1+(R/R_c)^2]^{-3\beta/2}$, with $\beta=1.4$ and $R_c=20$~kpc. Hence, 
though both populations are {\it more concentrated} towards the center, 
(iii) the relative contribution of [Fe/H]$<-2$ stars {\it increases} from 
$16\%$ in the inner halo (at Galactocentric distances $r<20$~kpc) to $>40\%$ 
in the outer halo, in good agreement with the observational results by Carollo 
et~al. (2007,2009)\footnote{The low-metallicity bias in their data sample, 
leading to an underestimate of the number of [Fe/H]$>-2$ stars, prevents a 
rigorous comparison between simulated and observed MDFs.}.
Our findings suggest that the outer halo between 
$20$~kpc~$\lsim r\lsim 40$~kpc is the most promising region to search 
for VMP stars, though it is obvioulsy harder to find more distant stars 
in magnitude limited surveys. 

The spatial distribution of very/extremely metal-poor stars is imprinted by 
two physical mechanisms: (i) the natural concentration of old stellar particles 
towards the more central region of the Galaxy; (ii) the gradual enrichment, 
through SN-driven outflows, of the MW environment out of which haloes virialize.

Our analysis predicts that the MDF of [Fe/H]$<-2$ stars weakly varies through 
the central $50$~kpc. Furthermore the all-radii MDF, including all the stars 
within $r_{vir}=239$~kpc, satisfactory matches the data once normalized to the 
number of observed stars. Two immediate consequences arise from these results: 
(i) the observed sample, which only explored a small portion of the Galactic 
halo, is a good representation of the complete one; (ii) the results of 
semi-analytical models, which cannot incorporate the spatial information, can 
nevertheless be compared with the observations to give reasonable approximate 
results.
\begin{figure}
  \centerline{\psfig{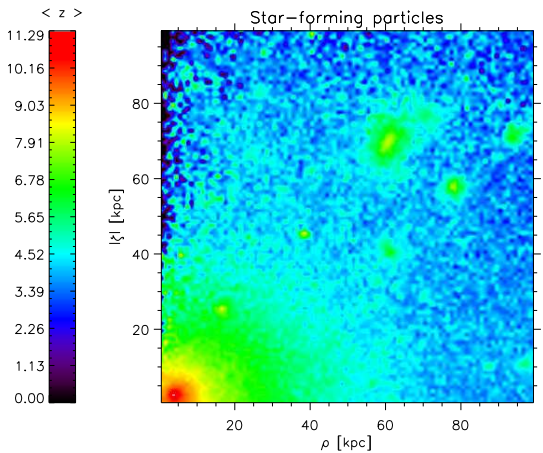}}
  \caption{Average formation redshift of DM particles hosting [Fe/H]$<-1$ stars
    in different regions of the $(\rho,|\zeta|)$ plane.}
\label{fig:4}
\end{figure} 

Beyond $60$~kpc the density of very metal-poor stars is maximum inside the 
``classical'' ($L>10^5 L_{\odot}$) dwarf galaxies. We find that 8 out of 10
host [Fe/H]$>-3$ stars only, in agreement with observations by Helmi et~al. 
(2006); in the remaining two extremely metal-poor stars represent a subdominant 
stellar population, making up only $\leq 13\%$ of the total stellar mass. 
Typically these galaxies virialize at $z\sim 6-8$, have a total mass 
$M_h = (1-7)\times 10^8\Msun$ (i.e. they are $<2\sigma$ fluctuations of the 
density field), and stellar mass $M_* = (0.5-7.5) \times 10^6 \Msun$. 
Interestingly, no ultra faint dwarf galaxies ($L<10^5L_{\odot}$) are found in 
our simulation, confirming that these newly-discovered satellites are probably 
left-overs of H$_2$ cooling mini-haloes (Salvadori \& Ferrara 2009), whose 
physics is not included in the present study.

We devote the final remark to our ``perfect mixing'' approximation. As the 
porosity increases rapidly with time ($Q\approx 1$ for $z\approx 7$), the MDFs 
from the inhomogeneous models are mostly consistent with those derived by using 
this approximation: the MDFs differences between the two mixing prescriptions 
(Fig.~1, left lower panel) are smaller than the $\pm 1\sigma$ error expected 
from averaging over different hierarchical merger histories (see Fig.~6 of 
SFS07). On the other hand, the true scatter in [Fe/H] at low metallicities and 
large radii may be substantially larger than in either of these models, as 
stars are strongly clustered towards the center of our simulation, which will 
reduce the true filling factor. The full physical modeling of metal mixing and 
diffusion remains one of the largest uncertainties in galaxy formation, and 
more work is required  before one can draw definite conclusions.
\section*{Acknowledgements}
We are grateful to A. Verdini for providing his carefully crafted IDL macros 
and to T.~Beers, D.~Carollo \& N.~Prantzos for careful reading of the draft 
and useful comments. 
\bibliographystyle{mn}
\bibliography{biblio}

\label{lastpage}

\end{document}